# Quantitative Comparison between AES and SIMS Depth Profiles of a Double Layer Structure of AlAs in GaAs Using the MRI-model.

S. Hofmann[a)], Max-Planck-Institute for Metals Research, Stuttgart, Germany
A.Rar, University of Alabama, Tuscaloosa, USA
D. W. Moon, Korea Research Institute of Standards and Science, Taejon, Korea
K. Yoshihara, National Research Institute for Metals, Tsukuba, Japan

## Abstract

Application of the so called mxing-roughness-information depth (MRI)-model to the quantitative reconstruction of the in-depth distribution of composition is demonstrated by comparing SIMS and AES depth profiles. A GaAs/AlAs reference sample consisting of two layers of AlAs (1 ML and 36 ML) separated by 44 ML of a GaAs matrix was depth profiled using almost identical sputtering conditions: $Ar^+$ ions of 3 keV impact energy and 52 (SIMS: CAMECA 4f) and 58 deg. (AES: VG Microlab 310F) incidence angle. Both the $Al^+$ intensity of the SIMS profile and the Al (LVV) intensity of the AES profile were quantified by fitting the measured profiles with those calculated with the MRI model, resulting in the same mixing length of 3.0 ± 0.3 nm, similar roughness parameter (1.4-2 nm), and negligible information depth (0.4 nm). Whereas practically no matrix effect was observed for AES as well as for $Al^+$ in the SIMS profile, quantification using dimer ($Al_2^+$) and trimer ($Al_3^+$) ions shows a marked nonlinearity between concentration and intensity, with the main effect caused by simple mass action law probability of cluster ion formation.

## Introduction

Depth profiling with ultra high resolution is a fast developing field of practical surface analysis[1-3]. Although the physical limits are almost attained, mainly by reducing the ion beam induced atomic mixing, for example by using low energy primary ions[4] or molecular cluster ions[2,5], the measured depth profile data still need some kind of deconvolution procedure to reconstruct the original in depth distribution of composition. Whereas noise in the data restricts precision, nonlinear relationships between signal intensity and/or sputtering time and elemental concentration and/or sputtered depth, respectively, impede usual deconvolution procedures and limit the accuracy of profile reconstruction[1]. The MRI (Mixing-Roughness-Information depth)-model[6,7] has been shown to be applicable for the reconstruction of the in- depth compositional distribution

[a)] electronic mail: hofmann@mf.mpi-stuttgart.mpg.de

from both AES and SIMS depth profiles. A stringent test of the validity of the MRI model requires the same result when using SIMS or AES on the same sample under identical sputtering conditions, where sputtering induced atomic mixing and surface roughness are expected to be the same. However, there is a basic difference in the analysis signal: (i) In contrast to SIMS, the elemental analysis signal in AES is independent of the primary ion flux, and is only little influenced by matrix effects. Therefore it is practically proportional to the elemental surface concentration. In SIMS, a non-linear dependence of the secondary ion current on surface composition is frequently observed[8], because the latter influences both sputter yield and ionisation probability; (ii) the information depth in AES depends on the Auger electron escape depth, i.e. their kinetic energy and emission angle, whereas in SIMS the information depth is dertemined by the escape depth of the secondary ions emitted from the first 1-2 atomic monolayers[8]; (iii) the dynamic range of the signal intensity is limited to about 2 - 3 orders of magnitude in AES, whereas the high dynamic range of SIMS typically covers about 6 orders of magnitude. Taking the above arguments into account, it is clear that SIMS is advantageous for profiling low concentration distributions such as dopant and delta layers, while AES is advantageous for interface and multilayer profiling.

In this paper, depth profile measurements by SIMS and AES are reported for a reference sample covering both regimes (delta layer and interface), where nonlinearities should be revealed. Quantitative profile reconstruction using the MRI model show direct similarity of the SIMS ($Al^+$) and AES (Al LVV) results and negligible nonlinearity. This is in accordance with earlier work[9] using dimer secondary ions ($Al_2^+$, $Ga_2^+$), where nonlinear behavior is expected[10]. There, a simple power law was shown to be sufficient for linearization. In this work, in addition trimer secondary ions ($Al_3^+$) intensity is shown to follow a similar power law, too.

## Experimental

The sample has the following GaAs/AlAs multilayer structure, (in ML, with 1 ML=0.28 nm) which was confirmed by high resolution TEM images:
40GaAs/1AlAs/44GaAs/36AlAs/GaAs(bulk)
Depth profiles were obtained with SIMS (CAMECA 4f) and AES (Microlab 310F) instruments under practically the same sputtering conditions. $Ar^+$ primary ions with 3 keV impact energy and an incidence angle of 52 deg (SIMS) and of 58 deg (AES) from the surface normal. The AES experiments were carried out with sample rotation.
In three different SIMS experiments, the following secondary ions were measured as a function of the sputtering time: $Al^+$, $Ga^+$, $Al^{++}$, $O^+$, $AlO^+$, $Al_2^+$, $Al_3^+$, $As^+$, $GaO^+$, and $Ga_2^+$. In this paper we will focus on $Al^+$ data, and compare $Al_2^+$ and $Al_3^+$ profiles and their quantitative evaluation with the MRI model.

In AES profiling, the Al LVV (68 eV) and Ga LMM (1063 eV) signals were recorded as a function of the sputtering time. The emission angle for Auger electrons was 40 deg from the surface normal, which gives the MRI - information depth parameter (taken here as mean electron escape depth) 1.58 nm for Ga LMM and 0.4 nm for Al LVV based on the attenuation length values from refs. [11,12]. Fortuitously, the latter value coincides with the one generally assumed for SIMS data (between 1 and 2 ML). Only the Al LVV profiles are considered here, the Ga LMM results were reported elsewhere[9].

## Results and Discussion

The experimental results were fitted with the MRI model results, as previously described (see for example refs.[6,7,13]). The model uses 3 physically well defined parameters: mixing zone length (w), roughness parameter (σ) and information depth (λ), which may be extracted from the measured depth profile or found independently[14], and calculates their influence on the shape of the sputter profile of an assumed layer structure. By comparison of the measured profile with the profile calculation, the optimum fit results in a quantitative reconstruction of the elemental in depth distribution of composition. Reference samples of the type used here serve to find the MRI parameters for the depth resolution functions that can be applied for quantification of unknown profiles[15].

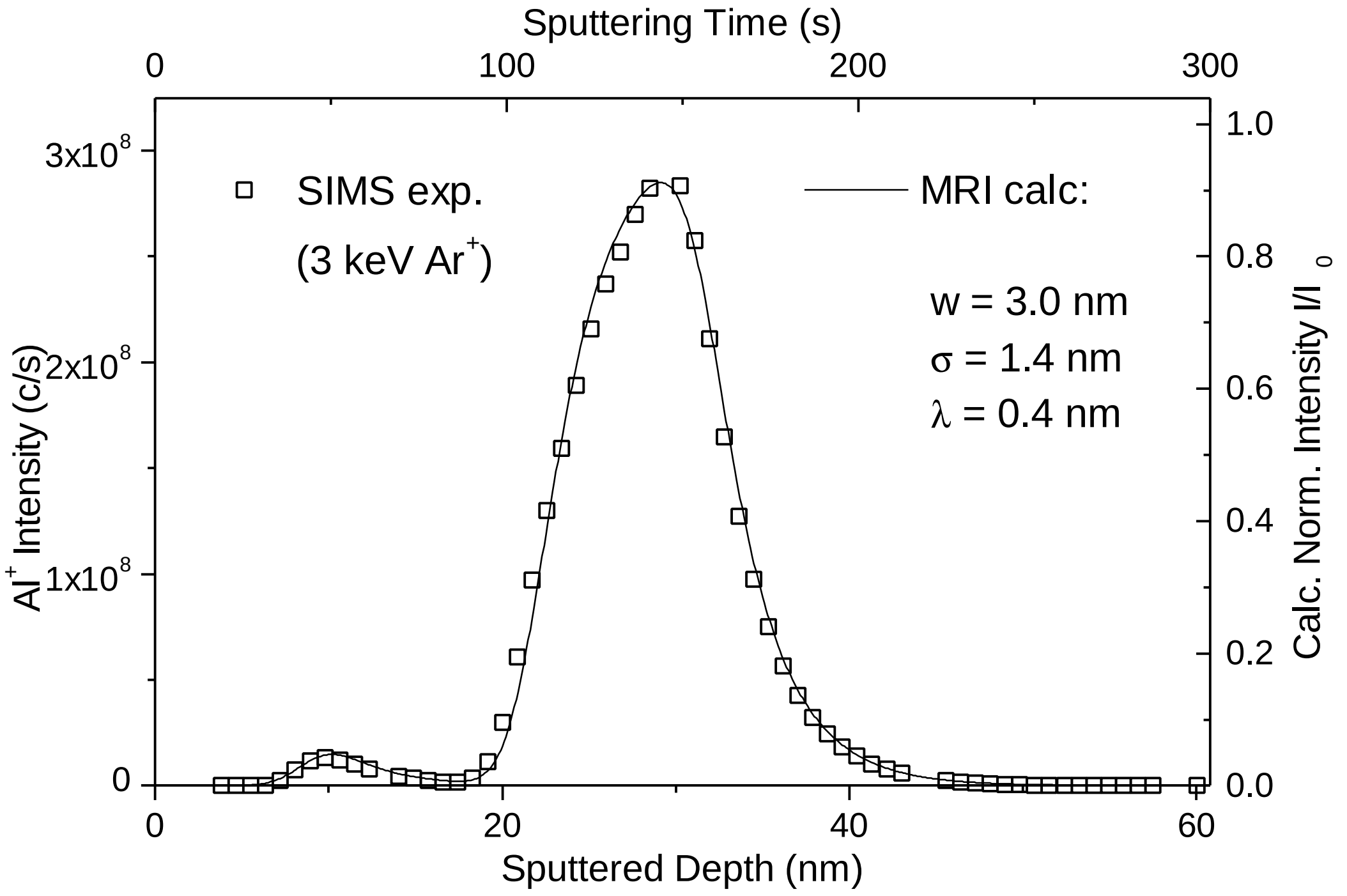

**Fig. 1a**

**Fig. 1a**. Linear plot of the SIMS depth profile obtained with 3 keV $Ar^+$ ions at 52 deg. incidence angle (open squares). The fit to the MRI calculation is shown as solid line. MRI- parameters are given in the inset.

The dependence of the $Al^+$ signal intensity on the sputtering time is shown in Fig. 1(a) together with the MRI fitting curve. The fit gives the length of the mixing zone, w = 3.0 nm, and the roughness, σ = 1.4 nm, together with the information depth for SIMS taken as λ = 0.4 nm (see above). The latter value is much lower than the other parameters, and its uncertainty has almost no influence on the experimental results. The value of the mixing length (w = 3.0 nm) compares reasonably well with the mean projected range of 3 keV Ar+ ions obtained from TRIM calculations ( GaAs: 2.3 ± 1.3 nm; AlAs: 2.5 ± 1.4 nm)[16]. Note that a reasonable fit was obtained by assuming a linear relationship for the delta layer (1 ML AlAs) and for the "thick" layer (36 ML AlAs) with the same parameters, i.e. over more than 3 orders of magnitude of concentration, as seen in the logarithmic plot in Fig.2a.

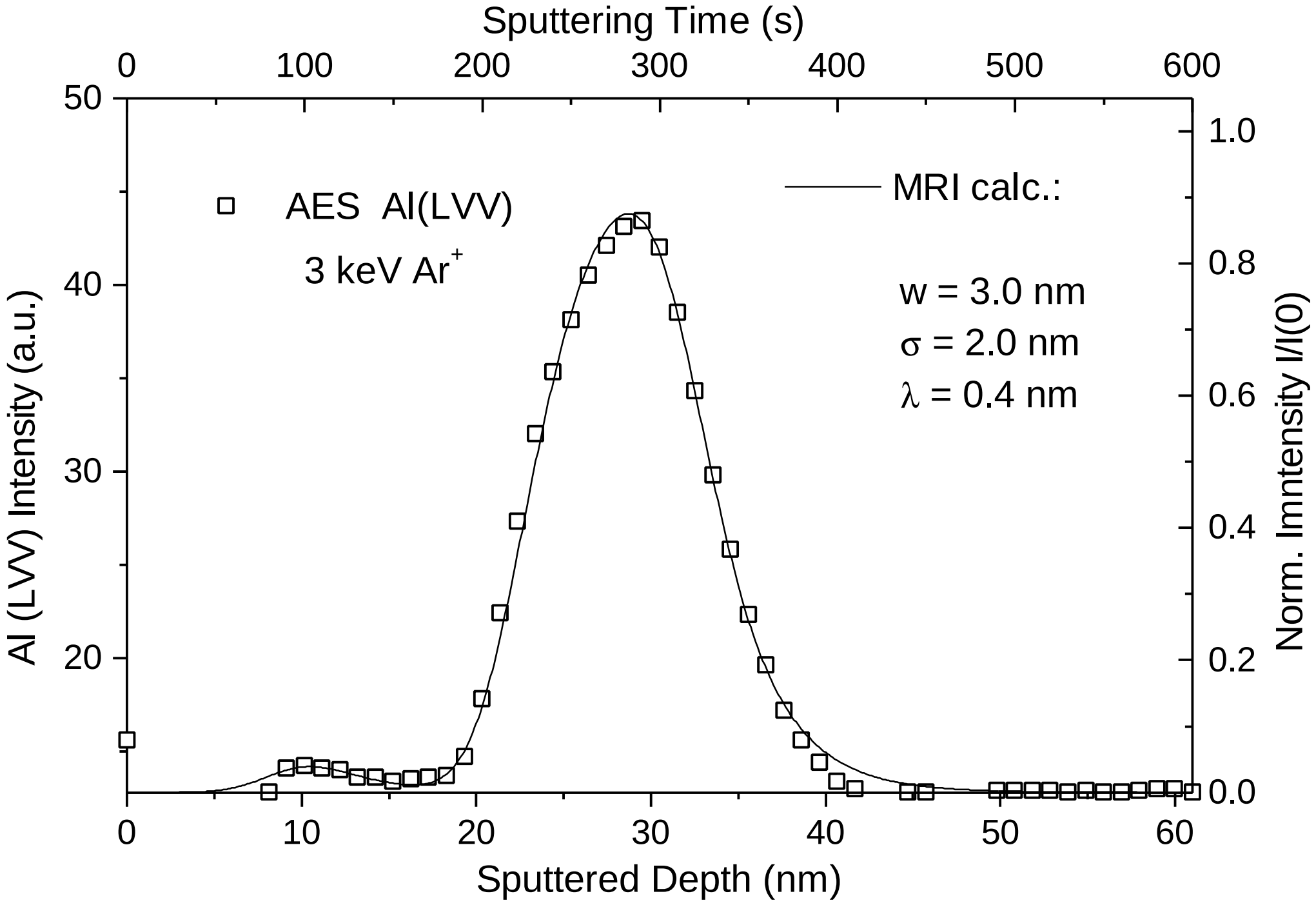

**Fig. 1b**

**Fig. 1b**. Linear plot of the AES depth profile obtained with 3 keV $Ar^+$ ions at 58 deg. incidence angle (open squares). The fit to the MRI calculation is shown as solid line. MRI- parameters are given in the inset.

AES depth profiling was carried out under practically the same experimental conditions in order to test the validity of data evaluation with the MRI model. The result is shown in Fig. 1(b). The mixing length, w = 3.0 nm, is indeed the same for SIMS and AES, within the error limit ( ± 0.2 nm ). This is expected because of the nearly identical sputtering conditions. The difference in the roughness parameter (σ = 2.0 nm for AES, 1.4 nm for SIMS) may be attributed to the slight difference in the ion incidence angle (AES:58 deg., SIMS: 52 deg.) different ion gun adjustments and to the use of sample rotation with only one analytical technique (AES).

Comparison of the logarithmic scale data of the AES profile in Fig. 2 (b) with Fig 2 (a) clearly demonstrates the disadvantage of AES at lower concentrations as compared to SIMS. The scatter of the AES data gets too large below 1% of the maximum signal of about 92% of the bulk intensity (see below).

*Quantification*:

Assuming a linear relationship between elemental signal intensity and concentration, the maximum intensity for a sufficiently thick layer of the respective element corresponds to 100 at% (or mole fraction = 1, with the corresponding intensity $I_0$).

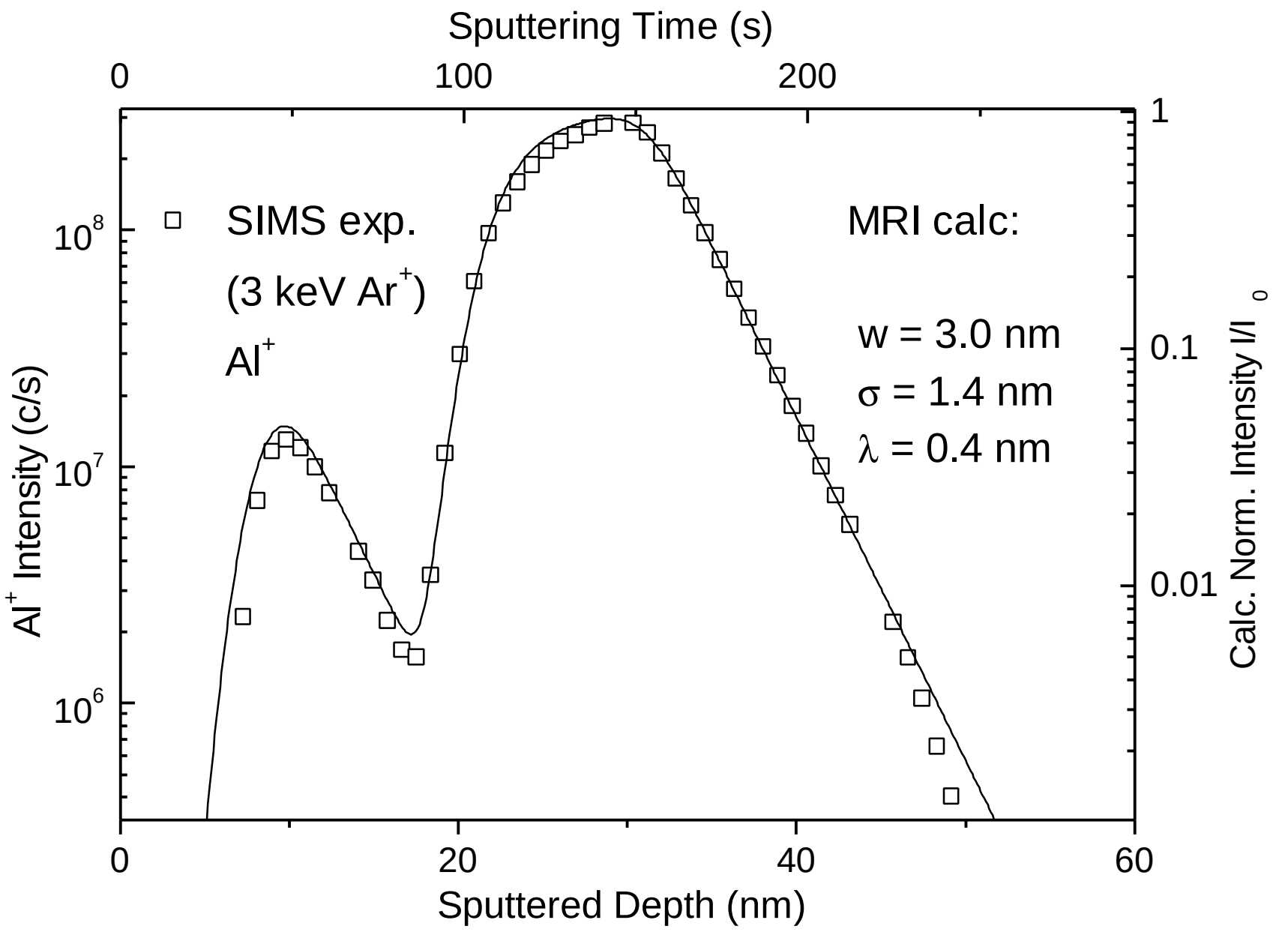

**Fig. 2a**

**Fig. 2a:** SIMS depth profile as in Fig.1, but plotted in logarithmic intensity scale to better disclose the low concentration delta layer profile on the left side.

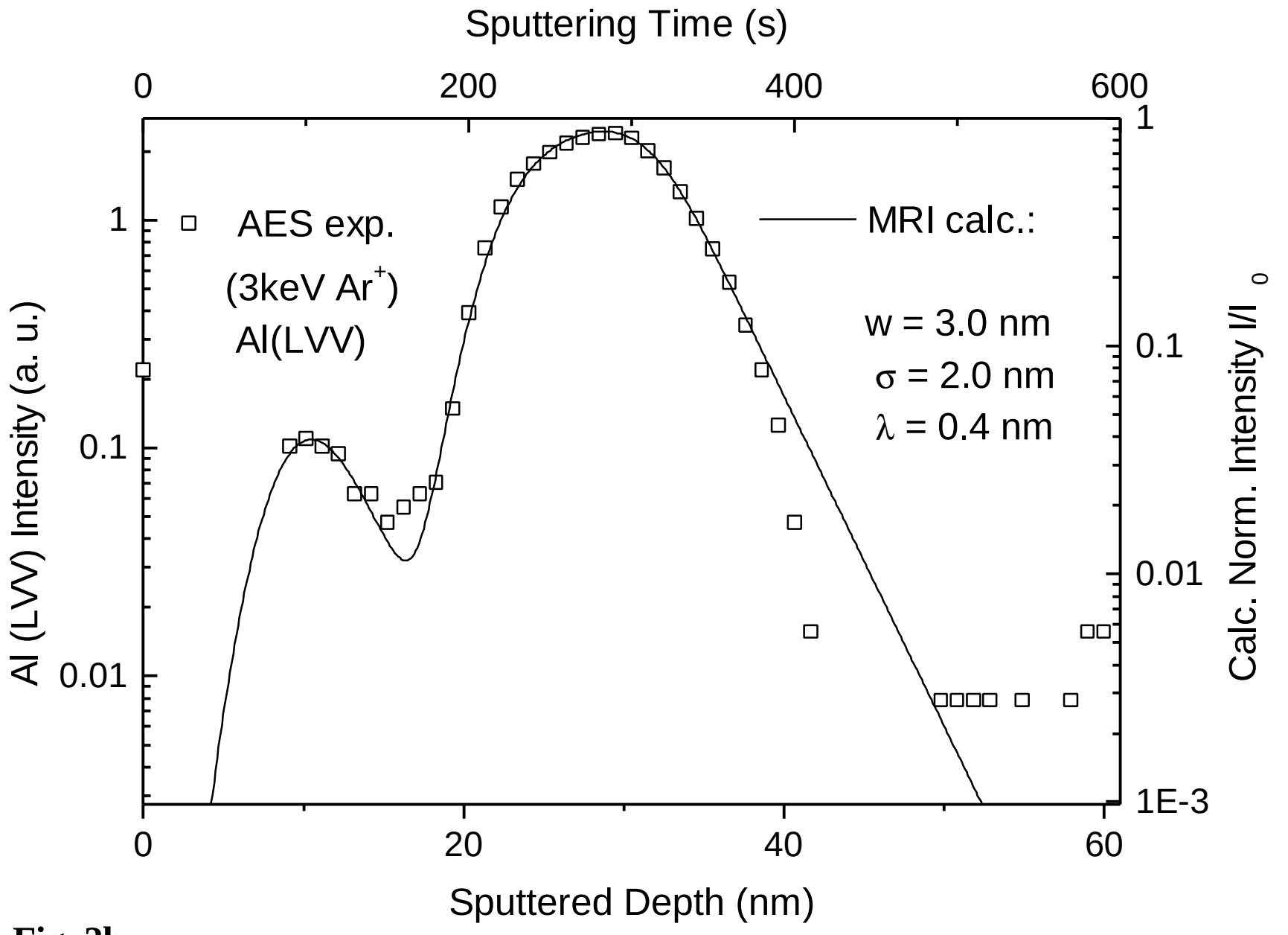

**Fig. 2b**

**Fig. 2b:** AES depth profile as in Fig.1, but plotted in logarithmic intensity scale to better disclose the low concentration delta layer profile on the left side. Note the limited dynamic range of AES.

The MRI model predicts the measured profile shape as a function of concentration of each monolayer and layer thickness. Therefore, if one point of the measured profile (usually the maximum intensity) is normalized to the calculated MRI-profile, all other profile points are quantified[17]. On the right hand side in Figs. 1a,b, and 2a,b, the scale is in fractions of the Al concentration in AlAs, which means that the peak signal of 36 ML of AlAs corresponds to 0.92 parts of Al in AlAs or to 46 at % Al. The maximum of the delta layer profile corresponds to about 0.05 parts or 2.5 at% of Al, respectively, if the original layer consists of exactly 1.0 ML of AlAs.

In the logarithmic SIMS data of Fig.2a, the solid line shows the MRI calculation with the given layer input as revealed from the TEM cross section image, i.e. 1 ML AlAs, 44 ML GaAs and 36 ML AlAs. The small deviation of less than 10% between the calculation and the measured points means that the upper limit of nonlinearity is 10%, a fairly good value for SIMS, which is confirmed by the completely independent AES measurements plotted on a logarithmic scale in Fig. 2b. Assuming constancy of the depth resolution function and that the only remaining nonlinearity is that of the relation between sputtering time and depth, i.e. a slightly different sputtering rate for GaAs and AlAs (< 5% as shown in ref. [6]), this could be implemented in the modified MRI model[15,18]. Because of the smallness of the effect, the sputtering rate differences can be neglected here. The maximum of the delta layer profile is calculated somewhat too high. Therefore, it is concluded that the layer is slightly thinner than 1 ML. With an Al content corresponding to 0.9 ML AlAs, the optimum fit is shown as solid line in Fig.3a. In addition, Fig. 3a shows the original (reconstructed) depth distribution (in ML: 40GaAs/0.9AlAs/44GaAs/36AlAs/GaAs...) together with the MRI quantification. To get an idea of the achieved accuracy, the calculated profiles for 0.7 ML and 1.1 ML AlAs are shown in Fig. 3b. It may be concluded that the error is less than ± 0.2 ML (±0.06 nm).

*Cluster ion emission*:
Cluster ion formation depends on many parameters and its quantitative description is difficult even for pure elements[10,19]. In a first order approximation, according to the mass action law, the reaction of cluster formation is expected to be proportional to the quadrature or to the 3rd power of the atomic concentration for dimers or trimers, respectively and we may expect for the measured intensities $I(Al_n^+)$ of those species:

$$I(Al^+) = k_1 * X(Al) \qquad (1)$$

$$I(Al_2^+) = k_2 * (X(Al))^2 \qquad (2)$$

$$I(Al_3^+) = k_3 * (X(Al))^3, \qquad (3)$$

where X(Al) is the mole fraction of Al and $k_1, k_2, k_3$ are factors which include the relative sensitivity factor for the respective ion species, matrix effects and experimental conditions. Assuming that these factors are constant, and because the information depth (i.e. mean escape depth of secondary ions) is small against the mixing length under the present conditions (3 keV Ar+ ions), the normalized intensity $I/I_0$, calculated with the MRI model, should be proportional to the instantaneous concentration, X(Al). (Note that "normalized intensity" means $I/I_0$ with $I_0$ the intensity for a sample with constant, known concentration independent of depth). Therefore the MRI model result can be directly compared with $I(Al^+)$ after eqn. (1), and with $I(Al_2^+)^{1/2}$ and $I(Al_3^+)^{1/3}$ after eqns. (2) and (3), if the assumptions and simplifications made above are valid. Of course similar relations are expected for Ga clusters, as discussed in ref.[9] for $Ga_2^+$ dimer ions.

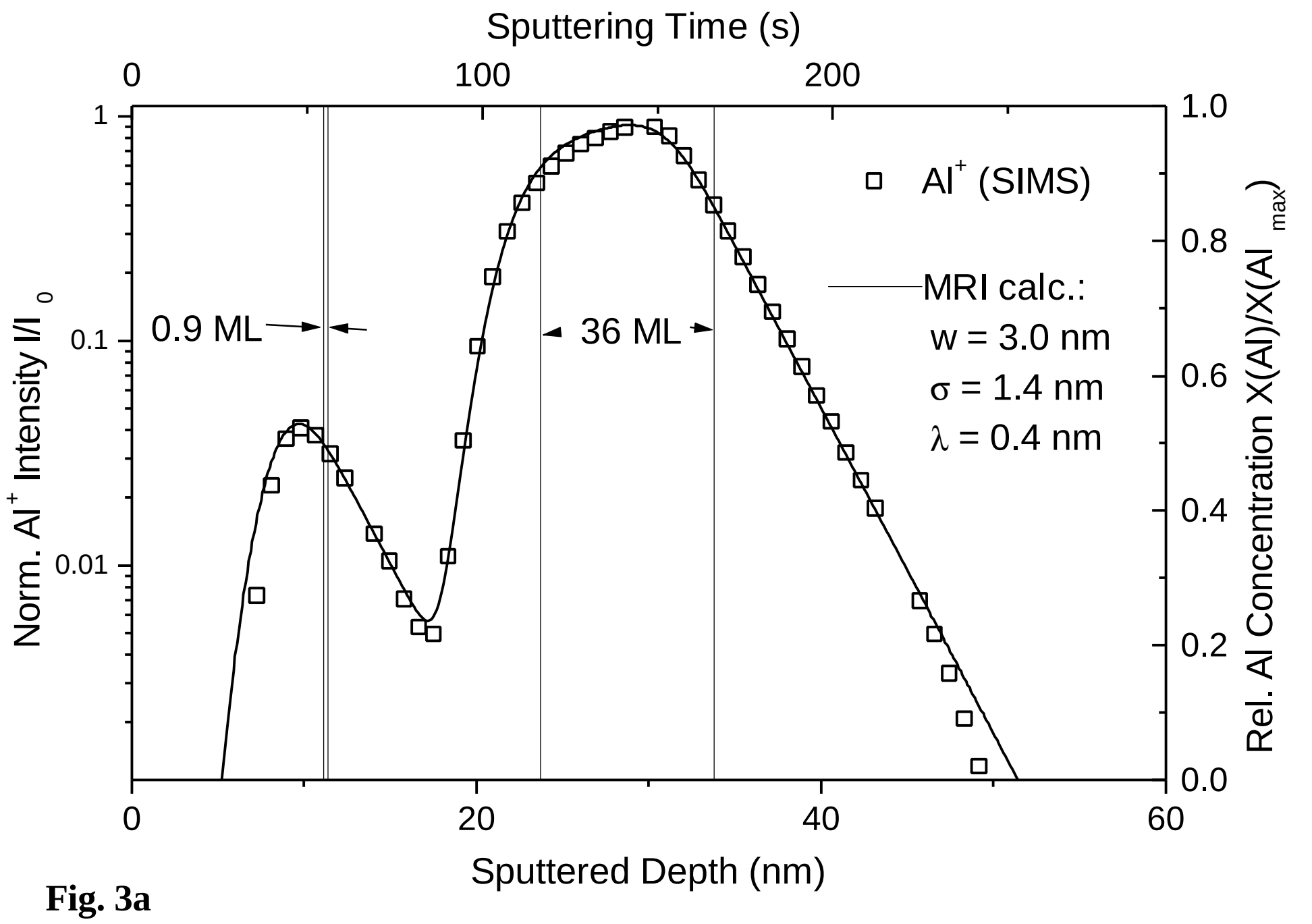

**Fig. 3a**

**Fig.3a**: Reconstructed , slightly modified original layer structure with the two layers of AlAs (0.9 and 36 ML) in GaAs, shown together with the experimental SIMS profile for $Al^+$ (open squares) and the optimized MRI calculation results.

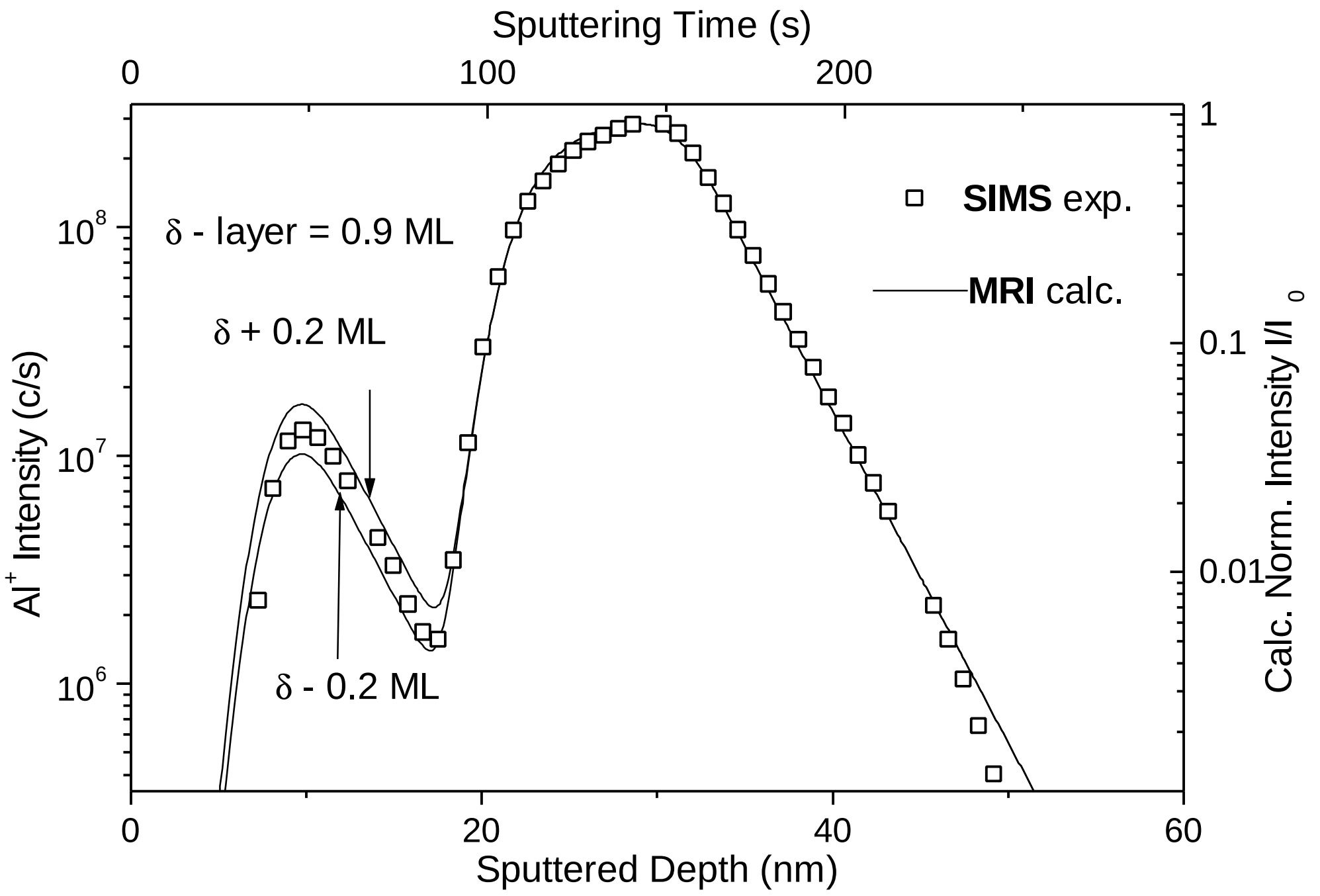

**Fig. 3b**

**Fig. 3b**: Visualization of the accuracy of profile calculation with the MRI model, which gives optimum fit for the delta layer consisting of 0.9 ML, as shown in Fig 2a, with an error of ± 0.2 ML of AlAs. Lines are calculations for 0.7 and 1.1 ML delta layer thickness.

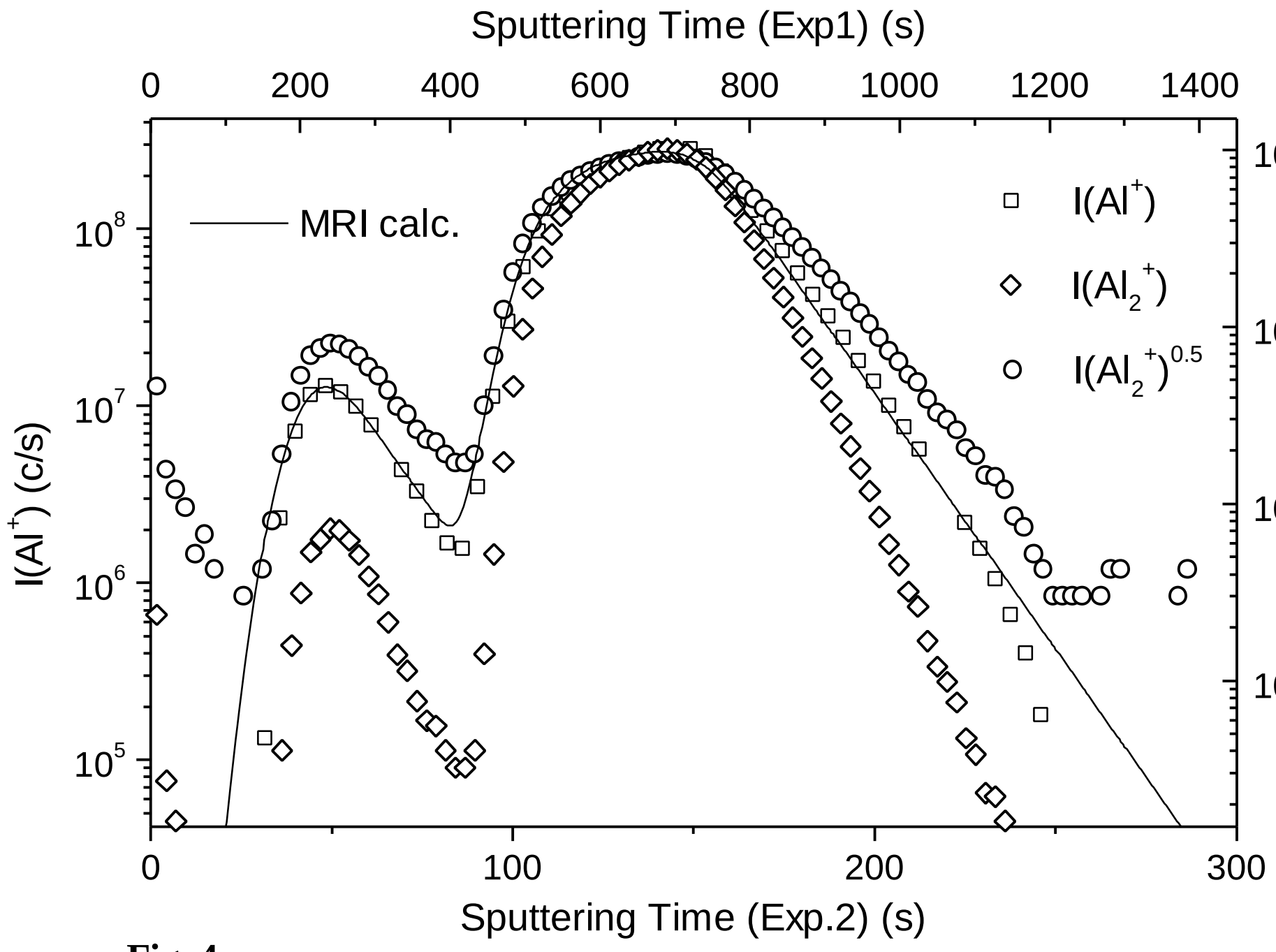

**Fig. 4a**

**Fig. 4a**: Comparison of the $Al^+$ intensity (fitted with MRI results) with the $Al_2^+$ intensity measured in another Experiment, and with the square root of that intensity according to eqn. (2). Note that $(Al_2^+)^{0.5}$ improves the fit, but the true exponent has to be between 0.5 and 1, and was empirically found to be 0.62.

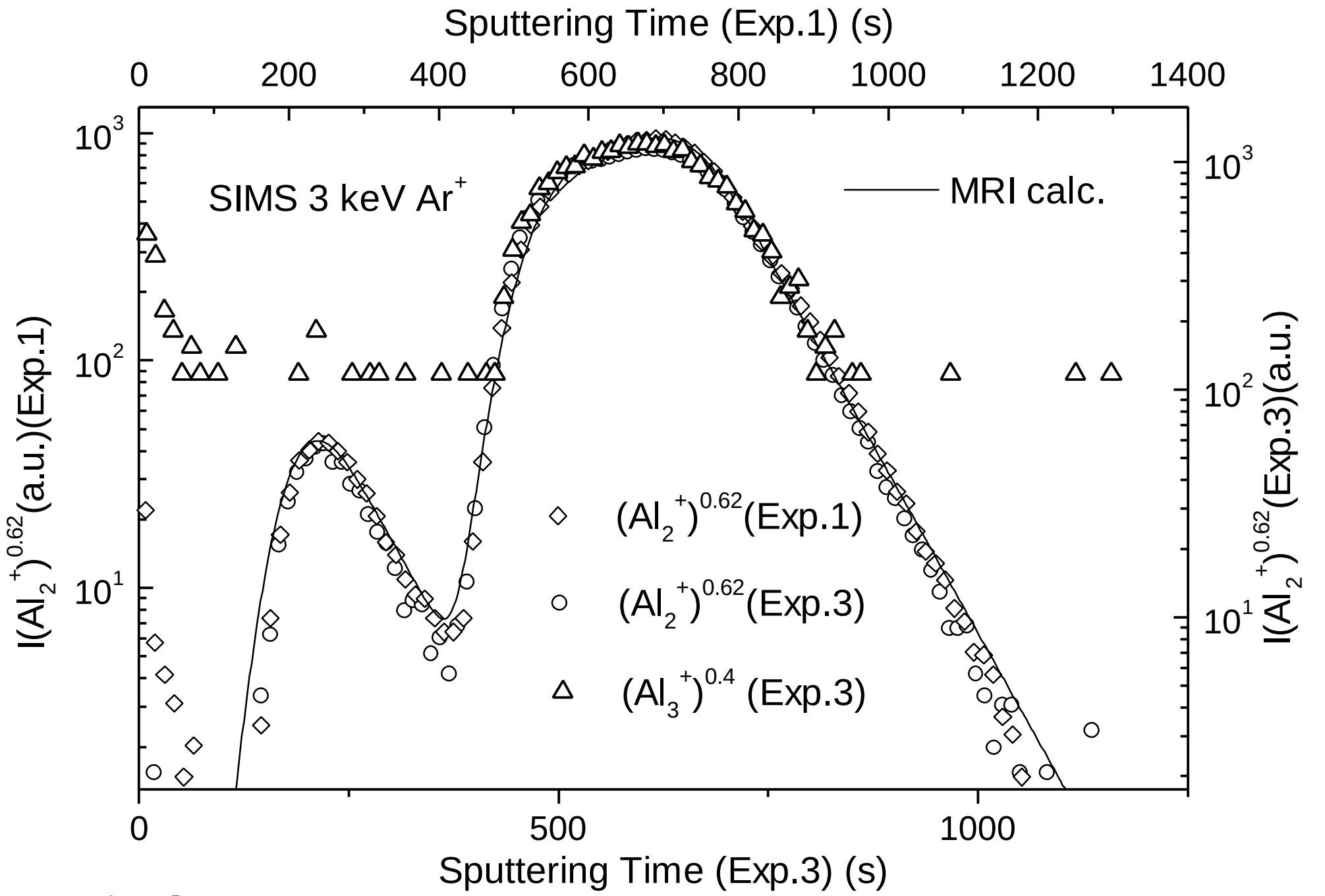

**Fig. 4b**

**Fig. 4b**: Optimum fit of the $Al_2^+$ cluster ion intensities (from 2 different experiments, Exp.1 and Exp.2), and that of the trimer $Al_3^+$, with the MRI calculation, by plotting $(Al_2^+)^{0.62}$ and $(Al_3^+)^{0.4}$ as a function of the sputtering time. Note that the optimum exponent for the trimer is again slightly higher (0.4) than the one expected from eqn. (3) (0.33).

Figs. 1a,b and 2a,b demonstrate the linear behavior expected after eqn.(1). Eqn. (2) predicts a nonlinear relation between $I(Al_2^+)$ and X(Al). Indeed, this is seen in the direct comparison of the $Al^+$ intensity (from Fig. 2a) with the $Al_2^+$ intensity normalized to the maximum value of each profile, as shown in Fig. 4a. According to eqns. (1) and (2), the square root of $I(Al_2^+)$, also shown in Fig.4a, should be equal to $I(Al^+)$. Indeed, the square root is more close to $I(Al^+)$, but the deviation is still significant and shows that the correct exponent should lie between 0.5 and 1. It is found to be 0.62 for the dimers, as already shown in ref. [1] by comparison with AES results. For $I(Al_3^+)$, the exponent after eqn. (3), 0.33, is likewise too small. Fig. 4b shows, that the dimer data, even from two independent experiments, and the trimer data can be linearized and consistently reproduced by the MRI model for the intesity exponent 0.62 for the dimer and 0.4 for the trimer. Note that there is more than an order of magnitude difference between the maximum count rates of $I(Al^+)$, $I(Al_2^+)$ and $I(Al_3^+)$. For the latter, the delta layer signal vanishes in the noise.

**Conclusions**

The direct comparison of SIMS and AES profiles obtained on the same samples of a bilayer structure of AlAs in GaAs with practically the same sputtering conditions provide a test for the applicability of the MRI model for profile quantification. Both the SIMS and AES data sets for Al can be fitted using the same parameter for the mixing length. While the information depth for Al is accidentally the same, the roughness is slightly different. However, this parameter is difficult to control, because it depends sensitively on the setup of the experiments, (e.g. sample rotation was only used in the AES profiling experiment). Even for the large concentration differences in the mixing zone between the delta layer profile and a thick layer profile, the intensity of $Al^+$ scales linearly with that calculated by the MRI model. In addition, depth profiles of the intensities of the secondary cluster ions $Al_2^+$ and $Al_3^+$ can also be fitted with the model by an empirical power law, $I(Al_2^+)^{0.62}$ and $I(Al_3^+)^{0.4}$, a small but significant modification of the exponents 0.5 and 0.33, respectively, expected under simplified conditions.